# An Efficient Encryption Scheme with Verifiable Outsourced Decryption in Mobile Cloud Computing

Jing Li[1], Zhitao Guan[1*], Xiaojiang Du[2], Zijian Zhang[3], Jun Wu[4]
1. School of Control and Computer Engineering, North China Electric Power University, China
2. Department of Computer and Information Science, Temple University, Philadelphia PA, USA
3. School of Computer, Beijing Institute of Technology, Beijing, China
4. College of Information Security Engineering, Shanghai Jiao Tong University, Shanghai, China
*Corresponding author: guan@ncepu.edu.cn

*Abstract*—With the increasing number of mobile applications and the popularity of cloud computing, the combination of these two techniques that named mobile cloud computing (MCC) attracts great attention in recent years. A promising public key encryption scheme, Attribute-Based Encryption (ABE), especially the Ciphertext Policy Attribute-Based Encryption (CP-ABE), has been used for realizing fine-grained access control on encrypted data stored in MCC. However, the computational overhead of encryption and decryption grow with the complexity of the access policy. Thus, maintaining data security as well as efficiency of data processing in MCC are important and challenging issues. In this paper, we propose an efficient encryption method based on CP-ABE, which can lower the overhead on data owners. To further reduce the decryption overhead on data receivers, we additionally propose a verifiable outsourced decryption scheme. By security analysis and performance evaluation, the proposed scheme is proved to be secure as well as efficient.

## I. INTRODUCTION

MCC is a service that allows resource constrained mobile users to adaptively adjust processing and storage capabilities by transparently partitioning and offloading the computationally intensive and storage demanding jobs on traditional cloud re-sources by providing ubiquitous wireless access [1]. Some of the researches on mobile cloud computing can refer to [2][3]. In order to ensure the confidential of data as well as the access control, lots of studies focus on combining mobile cloud computing and ABE (or its variants).

However, one of the drawbacks of ABE is that the computational overhead of encryption and decryption grows with the complexity of the access policy. To reduce the decryption computational overhead on DRs, some researchers consider that decrypt the ciphertexts by outsourced decryption cloud servers. What's more, these semi-trusted servers learn nothing about the messages and DRs can verify the correctness of the transformed ciphertexts. But the encryption overhead is not taken into account. To solve the problem, we propose an efficient encryption scheme based on CP-ABE, which can dramatically enhance data encryption efficiency without loss of data security and data privacy. Main contributions of this paper can be summarized as follows: 1) We propose an efficient encryption method, especially when users require to repeatedly encrypt the messages under the same access structure. 2) We additionally apply verifiable outsourced decryption to our scheme. 3) We show the security analysis and provide a detailed performance evaluation to demonstrate the advantages of our scheme.

The rest of this paper is organized as follows. Section 2 introduces the related work. In section 3, some preliminaries are given. In section 4, our scheme is stated. In section 5, security analysis is given. In Section 6, the performance of our scheme is evaluated. In Section 7, the paper is concluded.

## II. RELATED WORK

In [4], Sahai and Waters introduced attribute-based encryption (ABE) to achieve fine-grained encrypted access control. There must be at least *n* attributes overlapping between the ciphertext and user, the user could decrypt. In Goyal's construction [5], they extended the expressiveness of access structure by associating a ciphertext with a set of attributes, named Key-Policy Attribute Based Encryption (KP-ABE). In [6], Bethencourt et al introduced Ciphertext-Policy Attribute Based Encryption (CP-ABE), and used the attributes to describe a user's credentials. The data owners specified the access policy. The decryption keys consisted of a set of attributes without any tree structure.

In order to get a wider application, there are lots of researches on the combination of cloud and ABE. Wang et al [7] proposed a scheme to enable the confidential date to be stored on semi-trusted cloud servers. They realized the function by combining a HIBE system and a CP-ABE system. However, Yu et al [8] achieved fine-grainedness, scalability, and data confidentiality by exploiting and uniquely combining techniques of attribute-based encryption (ABE), proxy re-encryption, and lazy re-encryption. Some other applications of access control can be referenced in [9-12].

To reduce the computation overhead on users, some researchers tried to outsource the decryption to cloud. Green et al [13] proposed a scheme that allow users to outsource their ciphertexts to cloud, and the cloud servers could translate them into a (constant-size) El Gamal-style ciphertext, without disclosing any information about the user's messages. However, it couldn't verify the correctness of outsourced decryption. In [14], Lai et al guaranteed that a user can efficiently check if the transformation is done correctly without relying on any random oracles. Compared with this scheme, Lin et al [15] proposed a more efficient construction ABE with verifiable outsourced decryption, which reduced the bandwidth and the computation costs almost by half. Some other similar works can be seen in [16-18].

The similarity between the existing works and ours is that we are all based on ABE method, especially on outsourced decryption. However, in order to enhance the data processing efficiency, we propose a novel scheme that allow users not only to reduce the decryption computation overhead, but also the encryption computation overhead.

III. PRELIMINARIES

*A. Bilinear Maps*

Let $G_0$ and $G_1$ be two multiplicative cyclic groups of prime order $p$ and $g$ be the generator of $G_0$. The bilinear map $e$ is, $e: G_0 \times G_0 \to G_1$, for all $a, b \in \mathbb{Z}_p$:

- Bilinearity: $\forall u, v \in G_1, e(u^a, v^b) = e(u, v)^{ab}$
- Non-degeneracy: $e(g, g) \neq 1$
- Symmetric: $e(g^a, g^b) = e(g, g)^{ab} = e(g^b, g^a)$

*B. Complexity Assumptions*

**Definition 1** Discrete Logarithm (DL) Problem:

Let $G$ be a multiplicative cyclic group of prime order $p$ and $g$ be its generator. Given a tuple $<g, g^x>$, where $g \in_R G$ and $x \in \mathbb{Z}_p$ are chosen uniformly at random, as input, the DL problem is to recover $x$.

The DL assumption holds in $G$ is that no probabilistic polynomial-time (PPT) algorithm $\mathcal{A}$ can solve the DL problem with negligible advantage. We define the advantage of $\mathcal{A}$ as follows:

$$\Pr[\mathcal{A} < g, g^x > = x]$$

The probability is over the generator $g$, randomly chosen $x$ and the random bits consumed by $\mathcal{A}$.

*C. CP-ABE*

Secret Sharing Scheme (SSS) was proposed by Shamir [17] and Blakley [19] to share a secret among $n$ parties, only $n$ or more than $n$ parties can the secret be retrieved. This idea was already used to realize the tree-access structure. We use the definition adapted from [6]:

Let $\tilde{T}$ be an access tree, and the root node is denoted by $\mathcal{R}$. At the beginning of the encryption, we will conduct a polynomial for each node from top to bottom. Let $\tilde{S}$ be the set of leaf nodes value, where

$$\tilde{S}: \forall y \in S_{LeafNode}, \hat{C}_y = g^{q_y(0)}, \hat{C}_y' = H(att(y))^{q_y(0)}.$$

Finally, $\tilde{T}, \tilde{S}$ will be the components of ciphertext, denoting its corresponding access structure.

To retrieve the secret, we define the Lagrange coefficient $\Delta_{i,S}$ as follows:

For $i \in \mathbb{Z}_P$, and for $\forall x \in S$,

$$\Delta_{i,S(x)} = \prod_{j \in S, j \neq i} \frac{x-j}{i-j}$$

We now give the definition of security model for CP-ABE. In this security game, adversary is allowed to challenge on an encryption to an access structure $AS^*$ and query for any private key $SK$ such that $SK$ does not satisfy $AS^*$. The game proceeds as follows:

**Setup:** The challenger $\mathcal{C}$ runs this algorithm. It gives the public parameters PK to the adversary $\mathcal{A}$ and keeps MK to itself.

**Phase 1:** $\mathcal{A}$ issues queries for repeated private keys corresponding to sets of attributes $S_1, ... S_{q_1}$. ($q$ and $q_1$ is an integer that randomly chosen by $\mathcal{A}$ and $1 < q_1 < q$). If any of the sets $S_1, ... S_{q_1}$ satisfies the access structure $AS^*$, then aborts. Else, $\mathcal{C}$ generates the corresponding secret keys to the sets for $\mathcal{A}$. Then he submits a set of attribute and a ciphertext $CT$ and obtains the corresponding $M$ from $\mathcal{C}$.

**Challenge:** $\mathcal{A}$ submits two equal length messages $M_0$ and $M_1$ to $\mathcal{C}$. The challenger $\mathcal{C}$ randomly flips a coin $b$, and encrypts $M_b$ under the challenge access structure $AS^*$. Then the generated ciphertext $CT^*$ will be given to $\mathcal{A}$.

**Phase 2:** Repeat **Phase 1**, and the sets are turned from $S_1, ... S_{q_1}$ to $S_{q_1+1}, ... S_q$.

**Guess:** The adversary $\mathcal{A}$ outputs its guess $b' \in \{0,1\}$ for $b$ and wins the game if $b' = b$.

The advantage of an adversary $\mathcal{A}$ in this game is defined as

$$Adv(\mathcal{A}) = \left| \Pr[b' = b] - \frac{1}{2} \right|,$$

where the probability is taken over the random bits used by the challenger and the adversary.

**Definition 2** A CP-ABE scheme is CCA-secure if all polynomial time adversaries have at most a negligible advantage in the above game.

**Definition 3** We say that a CP-ABE scheme is CPA-secure if the adversaries cannot make decryption queries in Phase 1.

IV. DESCRIPTION OF OUR SYSTEM

*A. Setup*

The TA will run this setup algorithm and generate a set of public parameters. First, it chooses a bilinear group $G_0$ of prime order $p$ with generator $g$ and several random exponents: $\alpha, \beta, q \in \mathbb{Z}_p$. We introduce hash functions $H_m(), H_a()$ for plaintext and all of the attributes. The public key and the master key are published as:

$$\begin{aligned} PK &= \{G_0, g, h = g^\beta, e(g,g)^\alpha, g^q\} \\ MK &= \{\beta, g^\alpha, q\} \end{aligned} \quad (1)$$

*B. Key Generation*

**Key_Gen**(PK, MK, S) → SK

The set of attributes $S$ will be the input and a corresponding key is the output. This algorithm first chooses $r \in \mathbb{Z}_P$ and $r_j \in \mathbb{Z}_P$ at random for each attribute $j \in S$. Then it computes the key as:

$$SK = \left( D = g^{\frac{\alpha+r}{\beta}}, \forall j \in S, D_j = g^r H(j)^{r_j}, D_j' = g^{r_j} \right) \quad (2)$$

*C. Encryption*

In the traditional scheme [6], we will conduct an access tree for one access structure in each encryption process. For one access structure $AS_i$, its information will be contained in $\tilde{T}_i, \tilde{S}_i$ where,

$$\tilde{S}_i : \forall y \in S_{LeafNode}, \hat{C}_{i,y} = g^{q_{i,y}(0)}, \hat{C}_{i,y}' = H_a(att(y))^{q_{i,y}(0)} \quad (3)$$

The Data Owner (DO) can record the access structure $AS_i$ and consider it as an *Encryption Machine*. We define its *Encryption Machine* $EM_i$ as follows:

$$EM_i : \{\tilde{T}_i, \tilde{S}_i, s_i\} \quad (4)$$

Once DO requires to encrypt another new message $M$ with the $AS_i$, he will run this encryption algorithm.

**EM_Encrypt**($M$, $PK$, $MK$, $EM_i$)→$CT$

At first, this algorithm selects $u \in \mathbb{Z}_P, t \in \mathbb{Z}_P, \Delta s \in \mathbb{Z}_P$ at random. Then it will update and re-randomize the components in $EM_i$.

Set $s = s_i + \Delta s, q_y(0) = q_{i,y}(0) + \Delta s$ :

$$\tilde{C} = (M \| u)e(g,g)^{\alpha(s_i + \Delta s)t} = (M \| u)e(g,g)^{\alpha st} \quad (5)$$

$$C_1 = h^{(s_i + \Delta s)t} = h^{st} \quad (6)$$

$$C_0 = u^{H_m(M)q} \quad (7)$$

$$\tilde{S}_i' : \forall y \in S_{LeafNode}, \hat{C}_y = g^{(q_y(0)+\Delta s)t} = g^{q_y(0)t},$$
$$\hat{C}_y' = H_a(att(y))^{(q_y(0)+\Delta s)t} = H_a(att(y))^{q_y(0)t} \quad (8)$$

Finally, the ciphertext is published as:

$$CT = \left(\tilde{T}_i, \tilde{C}, C_0, C_1, \tilde{S}_i'\right) \quad (9)$$

### D. Decryption

The Date Receiver (DR) will run this algorithm to decrypt his ciphertext.

**Decrypt**($CT$, $SK$, $PK$)→$M$ or $\perp$

The decryption algorithm takes as input the $PK$, $SK$ and $CT = \left(\tilde{T}_i, \tilde{C}, C_0, C_1, \tilde{S}_i'\right)$. If the node $z$ is a leaf node, let $j=att(z)$, if $j \in S$, then,

$$F_z = \frac{e(D_j, \hat{C}_y)}{e(D_j', \hat{C}_y')}$$
$$= \frac{e(g^r H_a(j)^{rj}, g^{q_y(0)t})}{e(g^{rj}, H_a(att(z))^{q_z(0)t})} \quad (10)$$
$$= e(g,g)^{rtq_z(0)}$$

Let $x$ be the parent node of these nodes $z$, let $S_x$ be the set of its child nodes, the outputs will be stored in $S_x$. If $\forall z \in S_x, F_z = \perp$, the function returns $\perp$. Else, it can get the root node value by running recursive function:

$$F_x = \prod_{z \in S_x} F_z^{\Delta_{i,S_x'(0)}}, \text{ where } \begin{cases} n = index(z) \\ S_x' = \{index(z) : z \in S_x\} \end{cases}$$
$$= \prod_{z \in S_x} (e(g,g)^{rtq_z(0)})^{\Delta_{n,S_x'(0)}}$$
$$= \prod_{z \in S_x} (e(g,g)^{rtq_{parent(z)}(index(z))})^{\Delta_{i,S_x'(0)}} \quad (11)$$
$$= \prod_{z \in S_x} (e(g,g)^{rtq_x(n)})^{\Delta_{i,S_x'(0)}}$$
$$= e(g,g)^{rtq_x(0)}$$

If DR's attribute set satisfies $\tilde{T}_i$, it will get:

$$F_R = e(g,g)^{rts} \quad (12)$$

To recover $M$, it will perform the encryption as follows:

$$\frac{\tilde{C}F_R}{e(C_1, D)} = \frac{(M \| u)e(g,g)^{\alpha st} e(g,g)^{rst}}{e(h^{st}, g^{\frac{\alpha+r}{\beta}})} \quad (13)$$
$$= M \| u$$

### E. Verification

**Verify_M**($CT$, $M'$, $u'$, $PK$)→*True* or *False*

First, DR will utilize the Hash function and $u$ to calculate this $M$:

$$M', u' \to u'^{H_m(M')} \quad (14)$$

Verify that:

$$e(C_0, g) = e(u'^{H_v(M')}, g^q) \quad (15)$$

If (15) exits, this algorithm will output *True*. Else, it will output *False*.

### F. Outsource

To reduce the decryption overhead, we introduce the outsourced decryption into our scheme. This process requires several algorithms to complete. We will describe the details as follows:

**Gen_TK**($PK$, $SK$)→$TK$, $RK$

This algorithm takes as input the $PK$ and the user's $SK$, where

$$SK = \left(D = g^{\frac{\alpha+r}{\beta}}, \forall j \in S, D_j = g^r H(j)^{rj}, D_j' = g^{rj}\right).$$

It chooses $w \in \mathbb{Z}_P, v \in \mathbb{Z}_P$ uniformly at random. Then this algorithm re-randomize the $SK$ components with these two parameters.

The $TK$ is set as:

$$TK = \begin{pmatrix} D' = g^{\frac{\alpha+r}{\beta}} g^v = g^{\frac{\alpha+r}{\beta}+v}, \\ \forall j \in S, D_j' = g^r H(j)^{rj} g^w = g^{r+w} H(j)^{rj}, \\ D_j'' = g^{rj} \end{pmatrix} \quad (16)$$

The $RK$ is set as:

$$RK = \left(D_R = g^v, \hat{D}_R = g^{\frac{w}{\beta}}\right) \quad (17)$$

**Out_Decrypt**($CT$, $TK$, $PK$)→$T$ or $\perp$

This algorithm takes as input the $PK$, the $TK$ and a ciphertext, where

$$CT = \left(\tilde{T}_i, \tilde{C}, C_0, C_1, \tilde{S}_i'\right).$$

If the user's attribute set doesn't satisfy the access structure, it outputs $\perp$. Else, it computes as follows:

If the node $z$ is a leaf node, let $j=att(z)$, if $j \in S$, then,

$$F_z = \frac{e(D_j', \hat{C}_y)}{e(D_j'', \hat{C}_y')} = \frac{e(g^{r+w} H_a(j)^{rj}, g^{q_y(0)t})}{e(g^{rj}, H_a(att(z))^{q_z(0)t})} \quad (18)$$
$$= e(g,g)^{(r+w)tq_z(0)}$$

To obtain the root node value by running the following recursive function:

$$F_x = \prod_{z \in S_x} F_z^{\Delta_{i,S_x'}(0)}, \text{ where } \begin{cases} n = index(z) \\ S_x' = \{index(z) : z \in S_x\} \end{cases}$$

$$= \prod_{z \in S_x} (e(g,g)^{rtq_z(0)})^{\Delta_{n,S_x'}(0)}$$

$$= \prod_{z \in S_x} (e(g,g)^{rtq_{parent(z)}(index(z))})^{\Delta_{i,S_x'}(0)} \quad (19)$$

$$= \prod_{z \in S_x} (e(g,g)^{rtq_x(n)})^{\Delta_{i,S_x'}(0)}$$

$$= e(g,g)^{rtq_x(0)}$$

Finally, it will get:

$$F_R = e(g,g)^{(r+w)ts} \quad (20)$$

Then, it will perform the encryption as follows:

$$T = \frac{\tilde{C}F_R}{e(C_1, D')} = \frac{(M \| u)e(g,g)^{\alpha st}e(g,g)^{(r+w)st}}{e(h^{st}, g^{\frac{\alpha+r}{\beta}+v})}$$

$$= \frac{(M \| u)e(g,g)^{(\alpha+r+w)st}}{e(h^{st}, g^{\frac{\alpha+r}{\beta}+v})e(h^{st}, g^v)} \quad (21)$$

$$= \frac{(M \| u)e(g,g)^{(\alpha+r+w)st}}{e(g,g)^{(\alpha+r)st}e(h^{st}, g^v)^{\beta vst}}$$

This algorithm outputs $T$.

**DR_Decrypt**($CT$, $T$, $RK$, $PK$)$\to M$ or $\perp$

This algorithm takes as input a $CT = (\tilde{T}_i, \tilde{C}, C_0, C_1, \tilde{S}_i')$, $T$, $RK$, and $PK$. To recover $M$, it computes:

$$T \frac{e(C_1, D_R)}{e(C_1, \hat{D}_R)} = \frac{(M \| u)e(g,g)^{(\alpha+r+w)st}}{e(g,g)^{(\alpha+r)st}e(h^{st}, g^v)^{\beta vst}} \cdot \frac{e(h^{st}, g^v)}{e(h^{st}, g^{\frac{w}{\beta}})}$$

$$= \frac{(M \| u)e(g,g)^{(\alpha+r+w)st}e(g,g)^{\beta vst}}{e(g,g)^{(\alpha+r)st}e(h^{st}, g^v)^{\beta vst}e(g,g)^{wst}} \quad (22)$$

$$= M \| u$$

To verify the correctness of this message, DR can call the verification algorithm.

**Verify_M**($CT$, $M'$, $u'$, $PK$)$\to$ *True* or *False*

We set $M', u'$ to denote the message and the random number by outsource decryption. Then the algorithm computes:

$$M', u' \to u'^{H_m(M')} \quad (23)$$

Verify that:

$$e(C_0, g) = e(u'^{H_v(M')}, g^q) \quad (24)$$

If (24) exits, this algorithm will output *True*. Else, it will output *False*.

## V. SECURITY ANALYSIS

### A. System Security

**Theorem 1:** The security of our system is no weaker than that of [6].

**Proof:** We prove this theorem by the following game. Suppose that an adversary $\mathcal{A}$ can attack our scheme with non-negligible advantage.

**Setup:** The challenger $\mathcal{C}$ runs this algorithm. It gives the public parameters $PK = \{G_0, g, h = g^\beta, e(g,g)^\alpha, g^q\}$ to the adversary $\mathcal{A}$ and keeps $MK = \{\beta, g^\alpha, q, k\}$ to itself.

**Phase 1:** $\mathcal{A}$ issues queries for repeated private keys corresponding to sets of attributes $S_1, \ldots S_{q_1}$.

In addition, he also submits an access structure $AS^*$. If any of the sets satisfies the access structure $AS^*$, then aborts. $\mathcal{C}$ generates the corresponding secret keys $SKs$ to the sets for $\mathcal{A}$.

**Challenge:** $\mathcal{A}$ submits two equal length messages $M_0$ and $M_1$ to $\mathcal{C}$. The challenger $\mathcal{C}$ randomly flips a coin $b$, and encrypts $M_b$ under the challenge access structure $AS^*$. Then the generated ciphertext $CT^* = (\tilde{T}_i, \tilde{C}, C_0, C_1, \tilde{S}_i')$ will be given to $\mathcal{A}$.

**Phase 2:** $\mathcal{A}$ issues queries for repeated private keys as in Phase 1, and the sets are turned from $S_1, \ldots S_{q_1}$ to $S_{q_1+1}, \ldots S_q$.

**Guess:** The adversary $\mathcal{A}$ outputs its guess $b' \in \{0,1\}$ for $b$ and wins the game if $b' = b$.

Obviously, the game has properly simulated in that of [6]. Thus, if $\mathcal{A}$ can attack our scheme with non-negligible advantage, he can attack the CP-ABE [6] as well.

### B. Encryption

**Theorem 2:** The construction of ciphertext in our scheme is secure.

**Proof:** We prove this theorem by the following game. Suppose that an adversary $\mathcal{A}$ can attack the construction of ciphertexts in our scheme with non-negligible advantage. The game proceeds as follows:

**Setup:** The challenger $\mathcal{C}$ runs this algorithm. It gives the public parameters $PK = \{G_0, g, h = g^\beta, e(g,g)^\alpha, g^q\}$ to the adversary $\mathcal{A}$ and keeps $MK = \{\beta, g^\alpha, q, k\}$ to itself.

**Phase 1:** $\mathcal{A}$ issues queries for repeated private keys corresponding to sets of attributes $S_1, \ldots S_{q_1}$.

$\mathcal{C}$ generates the corresponding secret keys $SKs$ to the sets for $\mathcal{A}$.

**Challenge:** $\mathcal{A}$ submits an access structure $AS^*$ to $\mathcal{C}$ (any of the sets doesn't satisfy this structure). The challenger $\mathcal{C}$ randomly chooses a message and $u_1 \in \mathbb{Z}_P, u_2 \in \mathbb{Z}_P$, encrypts them under $AS^*$. Then he will sent the two different generated ciphertext $CT_1^*, CT_2^*$ (suppose that $s_2 = s_1 + \Delta s$), where

$$CT_1^* = (\tilde{T}_i, \tilde{C}_1, C_{1,0}, C_{1,1}, \tilde{S}_{i,1}'),$$
$$\tilde{C}_1 = (M \| u_1)e(g,g)^{\alpha s_1 t_1} = (M \| u_1)e(g,g)^{\alpha s_1 t_1},$$
$$C_{1,1} = h^{s_1 t_1},$$

$C_{1,0} = u_1^{H_m(M)q}$,

$\tilde{S}_{i,1}': \forall y \in S_{LeafNode}, \hat{C}_{1,y} = g^{q_{1,y}(0)t_1}, \hat{C}_{1,y}' = H_a(att(y))^{q_{1,y}(0)t_1}$,

and $CT_2^* = (\tilde{T}_i, \tilde{C}, C_0, C_1, \tilde{S}_i')$,

$\tilde{C}_2 = (M \| u_2)e(g,g)^{\alpha s_2 t_2} = (M \| u_2)e(g,g)^{\alpha s_2 t_2}$,

$C_{2,1} = h^{s_2 t_2}$,

$C_{2,0} = u_2^{H_m(M)q}$,

$\tilde{S}_{i,2}': \forall y \in S_{LeafNode}, \hat{C}_{2,y} = g^{q_{2,y}(0)t_2}, \hat{C}_{2,y}' = H_a(att(y))^{q_{2,y}(0)t_2}$.

**Phase 2:** $\mathcal{A}$ issues queries for repeated private keys as in Phase 1, and the sets are turned from $S_1, \ldots S_{q_1}$ to $S_{q_1+1}, \ldots S_q$.

**Output**: The adversary $\mathcal{A}$ outputs the *Encryption Machine* $EM_i^*: \{\tilde{T}_i, \tilde{S}_i, \Delta s\}$.

If adversary $\mathcal{A}$ wins the game, the challenger $\mathcal{C}$ obtains the $EM_i^*$, where $u_1 \neq u_2, t_1 \neq t_2, s_1 \neq s_2$. However,

$$y_1 = \frac{g^{q_{2,y}(0)t_2}}{g^{q_{1,y}(0)t_1}} = g^{q_{2,y}(0)t_2 - q_{1,y}(0)t_1}$$
$$= g^{x_1},$$
$$y_2 = \frac{H_a(att(y))^{q_{2,y}(0)t_2}}{H_a(att(y))^{q_{1,y}(0)t_1}} = H_a(att(y))^{q_{2,y}(0)t_2 - q_{1,y}(0)t_1}$$
$$= H_a(att(y))^{x_2},$$

To recover $x_1 = x_2 = q_{2,y}(0)t_2 - q_{1,y}(0)t_1$, he has to solve the DL problem. Since it is computationally infeasible to solve the problem in $G$. Thus, $\mathcal{A}$ cannot attack the construction of ciphertext in our scheme with non-negligible advantage.

### C. Verifiability

**Theorem 3:** The proposed construction of CP-ABE is verifiable.

**Proof:** The verifiability is described as a game between a challenger $\mathcal{C}$ and an adversary $\mathcal{A}$. In this security game, suppose that the adversary can attack the verifiability of our scheme with non-negligible advantage. The game proceeds as follows:

**Setup:** The challenger $\mathcal{C}$ runs this algorithm. It gives the public parameters $PK$ to the adversary $\mathcal{A}$ and keeps $MK$ to itself.

**Phase 1:** $\mathcal{A}$ issues queries for repeated private keys corresponding to sets of attributes $S_1, \ldots S_{q_1}$. If any of the sets satisfies the access structure $AS^*$, then aborts. Else, $\mathcal{C}$ generates the corresponding secret keys to the sets for $\mathcal{A}$. He submits a set of attribute and a ciphertext $CT$ and obtains the corresponding $M$ from $\mathcal{C}$.

**Challenge:** $\mathcal{A}$ submits a message $M^*$ and an access structure $AS^*$. The challenger $\mathcal{C}$ encrypts $M^*$ under the challenge access structure $AS^*$. Then the generated ciphertext $CT^*$ will be given to $\mathcal{A}$. Let $CT^* = (\tilde{T}_i, \tilde{C}, C_0, C_1, \tilde{S}_i')$, where $C_0 = \hat{u}^{H_m(M^*)q}$, $q$ is $MK$ and $\hat{u}$ is chosen by $\mathcal{C}$ randomly.

**Phase 2:** $\mathcal{A}$ issues queries for repeated private keys as in Phase 1.

**Output**: The adversary $\mathcal{A}$ outputs $M$ and $u$.

If he wins the game, the challenger $\mathcal{C}$ obtains $C_0' = u^{H_m(M)q}$, where $M \neq M^*$ and $u \neq \hat{u}$. Since $H_m$ is a collision-resistant hash function, with negligible probability, $H_m(M) = H_m(M^*)$.

### D. Outsource

**Theorem 4:** The security of our system with outsourced decryption is no weaker than that of [6].

**Proof:** We prove this theorem by the following game. Suppose that an adversary $\mathcal{A}$ can attack our scheme with outsourced decryption with non-negligible advantage.

**Setup:** The challenger $\mathcal{C}$ runs this algorithm. It gives the public parameters $PK = \{G_0, g, h = g^\beta, e(g,g)^\alpha, g^q\}$ to the adversary $\mathcal{A}$ and keeps $MK = \{\beta, g^\alpha, q, k\}$ to itself.

**Phase 1:** $\mathcal{A}$ issues queries for repeated private $SK$ and $TK$ keys corresponding to sets of attributes $S_1, \ldots S_{q_1}$.

$\mathcal{C}$ generates the corresponding secret keys $SKs$ and $TKs$ for $\mathcal{A}$. The $RKs$ are unknown to $\mathcal{A}$.

**Challenge:** $\mathcal{A}$ submits two equal length messages $M_0$ and $M_1$ and an access structure $AS^*$ to $\mathcal{C}$. The challenger $\mathcal{C}$ randomly flips a coin $b$, and encrypts $M_b$ under the challenge access structure $AS^*$ (the sets above cannot satisfy this access structure). Then the generated ciphertext $CT^*$ will be given to $\mathcal{A}$.

**Phase 2:** $\mathcal{A}$ issues queries for repeated private keys as in Phase 1, and the sets are turned from $S_1, \ldots S_{q_1}$ to $S_{q_1+1}, \ldots S_q$.

**Guess:** The adversary $\mathcal{A}$ outputs its guess $b' \in \{0,1\}$ for $b$ and wins the game if $b' = b$.

Obviously, the game has properly simulated in that of [6]. Thus, if $\mathcal{A}$ can attack the construction of outsourced decryption in our scheme with non-negligible advantage, he can attack the CP-ABE [6] as well.

## VI. PERFORMANCE EVALUATION

To evaluate the performance of our system, we implemented a testing environment with the help of the cpabe-toolkit [20].

We encrypt the message under an access tree that contained ten levels and a hundred leaf nodes. As shown in Fig.1 (a), the average encryption time is greatly reduced.

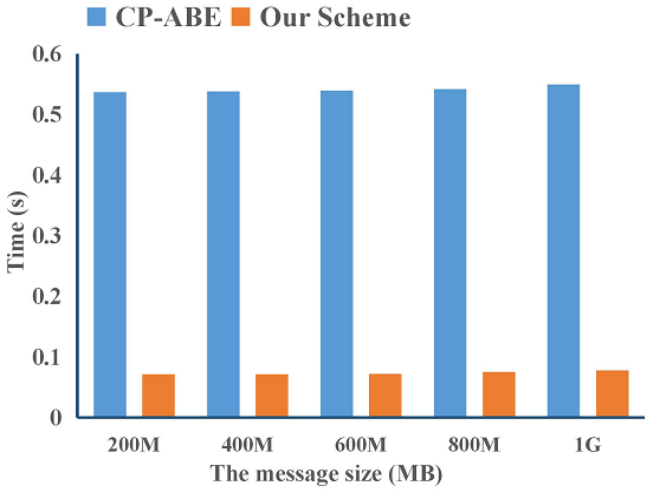

Fig. 1(a). Comparison of the average encryption time between the tradition scheme in [6] and ours, when the size of message grows.

Additionally, we select the 1GB message and encrypt it under the above tree for different times. As shown in Fig.1 (b), along with the number of encryption times grows, the average encryption time of CP-ABE is approximately flat, while the time in our scheme is slowly decreasing.

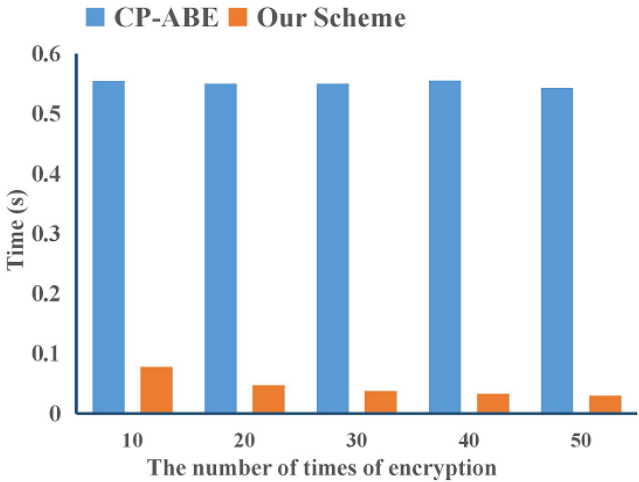

Fig. 1(b). Comparison of the average encryption time between the tradition scheme in [6] and ours, when the number of encryption times grows.

## VII. CONCLUSION

Cloud computing is an emerging paradigm, ABE is widely used for realizing access control in cloud. Generally, users encrypt their data in local and store them into cloud. However, maintaining data security and system efficiency meanwhile is a challenging issue. Existing related schemes seldom take efficiency of data encryption into consideration. To solve this problem, this paper proposes an efficient encryption scheme based on CP-ABE, which can not only guarantee secure data access, but also reduce overhead both on DO and DR. The security analysis shows that the proposed scheme can meet the security requirement. The evaluations show the advantages on the efficiency of data encryption.

ACKNOWLEDGMENT

This work is partially supported by Natural Science Foundation of China under grant 61402171, the Fundamental Research Funds for the Central Universities under grant 2016MS29, as well as by the US National Science Foundation under grant CNS-1564128.